\documentclass[useAMS,usenatbib,usegraphicx]{mn2e}
\usepackage{subfigure}      

\title[]{Planet formation: Statistics of spin rates and obliquities of extrasolar planets}
\author[Y. Miguel and A. Brunini]{Y. Miguel$^{1,2}$\thanks{E-mail: ymiguel@fcaglp.unlp.edu.ar} and A. Brunini$^{1,2}$\thanks{Member of the Carrera del Investigador Cient\'\i fico. Consejo Nacional de Investigaciones Cient\'\i ficas y T\'ecnicas (CONICET).E-mail: abrunini@fcaglp.unlp.edu.ar}\\
$^1$Facultad de Ciencias Astron\'omicas y Geof\'\i sicas. Universidad
Nacional de La Plata. Paseo del Bosque s/n, La Plata (1900), Argentina.\\
$^2$Instituto de Astrof\'\i sica de La Plata (CCT La Plata-CONICET, UNLP), Paseo del Bosque s/n, La Plata (1900), Argentina}

\begin{document}  

\pagerange{\pageref{firstpage}--\pageref{lastpage}}

\label{firstpage}

\maketitle
        
\begin{abstract}
  
We develop a simple model of planetary formation, focusing our attention on those planets with masses less than $10 M_{\oplus}$ and studying particularly the primordial spin parameters of planets resulting from the accretion of planetesimals and produced by the collisions between the embryos. As initial conditions, we adopt the oligarchic growth regime of protoplanets in a disc where several embryos are allowed to form. We take different initial planetary system parameters and for each initial condition, we consider an evolution of $2x10^7$ $years$ of the system. We perform simulations for $1000$ different discs, and from their results we derive the statistical properties of the assembled planets. We have taken special attention to the planetary obliquities and rotation periods, such as the information obtained from the mass and semi major axis diagram, which reflects the process of planetary formation. The distribution of obliquities was found to be isotropic, which means that planets can rotate in direct or indirect sense, regardless of their mass. Our results regarding the primordial rotation periods show that they are dependent on the region where the embryo was formed and evolved. According to our results, most of the planets have rotation periods between $10$ and $10000 \; hours$ and there are also a large population of planets similar to terrestrial planets in the Solar System.

\end{abstract}

\begin{keywords} 
Planets and satellites: formation\ - Solar System: formation\
\end{keywords}

\section{Introduction}

 Following the first discovery of an extrasolar planet around 51 Peg \citep{b20}, the number of exoplanets known has risen to 429. Although most of them are giant planets, the improvements in observational techniques have ensured that planets with masses less than $15 M_{\oplus}$ have started being detected with radial velocity survey (e.g., \citet{b22,b23,b24,b25,b21}) and gravitational microlensing survey \citep{b26}. 

Although most of extrasolar planets so far discovered are giant planets, several statistical models for planetary growth presented in the last years suggest that a large number of small planets who fail to have enough mass to start the gas accretion onto the core exists \citep{b9,b2,b28}, and has still not been able to be discovered \citep{b27}. At the time, several projects are in progress to detect terrestrial planets, we expect that they may find more Earth-size planets in  a close-future, but today, the sample is not enough and we also have to rely on what we know from our own Solar System, and through computational models of planetary formation.

This evidence supports the standard scenario, where terrestrial planets are formed through the next different stages: 1) agglomeration of dust particles through physical collisions and setting in the protoplanetary disc, 2) planetesimal formation from grains in a thin midplane \citep{b31,b32}, 3) runaway (e.g., \citet{b33}) and oligarchic \citep{b4,b15} accumulation of planetesimals to form protoplanets and 4) giant impact stage, where the embryos formed by oligarchic growth collide with one another to form planets \citep{b30}. 

The final stage of terrestrial planetary formation is the particular importance as it has a deep effect on the final characteristics of the planets: mass, orbital and spin parameters. After this stage of planetary formation, the spin parameters of the planets change and evolve due mainly to tidal interactions with their satellite and host star. All of the terrestrial planets in our Solar System do not maintain their primordial spin state and this is the reason why we unknown what primordial planetary spin would be expected to find in a protoplanet. So questions as, what are the typical obliquity and rotation period that characterise the primordial planets? and how many collisions suffers a planetary embryo along its firsts years of formation? remain uncertain.    

A few works dealing with the study of planetary spins have been presented. \citet{b34} have examined the accretion rate of spin angular momentum by a planet immersed in a differentially rotating disc of planetesimals. They determined the mass and spin accreted by the embryos as a function of the velocity dispersion of the disc particles and the ratio of the planetary radius to the Hill radius. They found that if a protoplanet grows by accreting a large number of small planetesimals the spin angular momentum of the planet will be determined by the called ``ordered component'', but if a few giant impacts occur, most of the spin will be contributed by the ``stochastic component''. \citet{b35} have investigated the spin of a planet which accreted in a disc of planetesimals with non uniform spatial distribution. They results show that the ordered component can dominate the final spin of the planet only if half of the size of the planet was acquire by the accretion of small planetesimals and the size of the impactors is not too large.      

On the other hand \citet{b13} and \citet{b19} have studied through N-body simulations the last stages of the terrestrial planet formation. They analysed the planetary obliquities as those found only considering the impacts between large embryos and have shown that this obliquities are expected to be represented by an isotropic distribution, result that was confirmed and generalized by \citet{b14}, who also considered an N-body code, but analysed a larger sample of embryos considering the standard disc model. 

Our principal aim is to make a statistical study of the primordial spin parameters of planets (obliquity and rotation period), resulting from the accretion of planetesimals and also due to the collisions between the emerging embryos. To this end we take different initial conditions, meaning different discs, stars, initial number of embryos, and study the primordial planetary spins in different systems with the intention of obtain a better understanding of what we should expect to find in the Universe. We also analyse what are the consequences of planetary impacts in the mass and semi major axis diagram, considering embryos with masses less than $10 M_{\oplus}$. Our semi-analytical model takes as initial condition the oligarchic growth regime of protoplanets and allows them to migrate, fact that has a huge influence on the number of collisions suffered by an embryo. We adopt a perfect accretion in collisions, supposition that was also considered by other authors \citep{b13,b19,b14}, but which says that the results should be interpreted cautiously.   

Each one of the 1000 systems considered, evolves for $2 x 10^7$ $years$ and we analyse the results statistically, finding an isotropic distribution of obliquities and where most of the planets rotate with a period between $10$ and $10000 \; hours$. We also found a large population of planets with the characteristics of terrestrial planets in the Solar System.

\section{Model and basic equations}

In this section we explain the model and basic equations consider in the work. As we take special attention to the planetary spin, the model adopted for the acquisition of angular momentum due to accretion and collisions between the embryos will be explain in detail. On the other hand the model for planetary growth and orbital evolution is essentially the same developed in our previous works \citet{b1,b2}, which is a very simple model based on the oligarchic growth regime and consider type I and II migration. For the sake of completeness, we will summarize it briefly below.

\subsection{Planetary growth}

We consider a protoplanetary nebula structure based on the minimum mass solar nebula (MMSN) \citep{b3}, where the surface density of solids at a distance $a$ from the central star is

\begin{equation}\label{distrisolidos}
\Sigma_d=7 f_d \eta_{ice} \big(\frac{a}{1au}\big)^{-\frac{3}{2}} gcm^{-2}
\end{equation}
with  $\eta_{ice}$ a step-function which takes the value $1$ inside the ice condensation radius and $4$ outside it, expressing the effect of water ice formation. The snow line is located at $a_{ice}=2.7\big(\frac{M_{\star}}{M_{\odot}}\big)^2$ $au$ from the central star of mass $M_{\star}$. 

 On the other hand the volume density of gas is

\begin{equation}
\rho_{gas}(a,z)=\rho_{g,0}(a)e^{\frac{z^2}{h(a)^2}}  gcm^{-3}
\end{equation}
 where $\rho_{g,0}(a)=1.4x10^{-9} f_g \big(\frac{a}{1au}\big)^{-\frac{11}{4}}  gcm^{-3}$. 

The parameters $f_d$ and $f_g$ state the solid and gas mass in the disc in terms of the MMSN model. We consider a large population of discs (1000 in each simulation), where we assume that $f_g$ follows a Gaussian distribution in terms of $log_{10}f_g$, centered at $0$, with dispersion of 1 and $f_d$ is taken as $f_d=f_g10^{0.1}$ in order to consider more metallic discs. 
 
Both discs are not time-invariant. The gaseous disc change globally, decaying exponentially with a characteristic time-scale of $\tau_{disc}$, which takes values between $10^6-10^7 years$ in accordance to current estimates of disc lifetimes around young solar type stars \citet{b16} and the solid disc change locally, suffering the depletion of planetesimals produced by the effect of core's accretion. The disc of planetesimals also interacts with the nebular gas, this gas drag effect cause a radial motion of planetesimals before they become large enough to decouple from the disc gas \citep{b6,b7}, we also consider this effect which was explained in detail in our previous work \citep{b2}. 

The protoplanetary discs are extended between $a_{in}\simeq 0.0344 2 \big(\frac{M_{\star}}{M_{\odot}}\big)^2 au$ \citep{b5} and $30au$. The first initial core is located at $a=a_{in}$, the rest of the cores are separated $10r_H$ each other until the end of the disc is reached. Their initial masses are given by the minimum mass necessary for starting the oligarchic growth stage \citep{b4,b15}, 

\begin{equation}\label{Masa-inicial}
M_{oli}\simeq \frac{1.6 a^{\frac{6}{5}}10^{\frac{3}{5}}m^{\frac{3}{5}}\Sigma_d^{\frac{3}{5}}}{M_{\star}^{\frac{1}{5}}}
\end{equation}
with m the effective planetesimal mass.  

The solid accretion rate for a core in the oligarchic growth regime, considering the particle-in-a-box approximation \citep{b8} is

\begin{equation}
\frac{dM_c}{dt}=10.53 \Sigma_d \, \Omega \, R_p^2\bigg(1+\frac{2GM_t}{R_p\sigma}\bigg)
\end{equation}
where $\Omega$ is the Kepler frequency, $R_p$ and $M_t$ are the planet's radius and total mass (solid and gas) and $\sigma$ is the velocity dispersion which depends on the eccentricity of the planetesimals in the disc. \citet{b7} obtain an expression for the rms eccentricity of the planetesimals when gravitational perturbation of the protoplanets are balanced by the dissipation due to the gas drag, which is   

\begin{equation}\label{em}
e_m^{eq}=\frac{1.7\, m^{1/15} \, M_t^{1/3} \, \rho_m^{2/15}}{b^{1/5}\, C_D^{1/5}\, \rho_{g,0}^{1/5}\, M_{\star}^{1/3}\, a^{1/5}}
\end{equation}
where $b$ is the orbital separation between the embryos in Hill radius units, ($b=10$), $C_D$ is a dimensionless drag coefficient which is $\simeq 1$ and $\rho_m$ is the planetesimal bulk density. With this expression they found the next oligarchic-regime growth rate which includes the evolution of the planetesimal rms $e$ and $i$,

\begin{equation}\label{core-accretion}
\frac{dM_c}{dt}\simeq \frac{3.9 b^{\frac{2}{5}}C_D^{\frac{2}{5}}G^{\frac{1}{2}}M_{\star}^{\frac{1}{6}}\rho_{gas}^{\frac{2}{5}}\Sigma_d}{\rho_m^{\frac{4}{15}}\rho_M^{\frac{1}{3}}a^{\frac{1}{10}}m^{\frac{2}{15}}}M_t^\frac{2}{3}
\end{equation}
where $\rho_M$ is the embryo bulk density, which is equal to the planetesimals density, then hereafter $\rho_M=\rho_m=\rho$.

The growth of the cores terminate when the solid surface density in their feeding zones is zero, which is caused by a combination of these factors: the embryos consume planetesimals on their feeding zones, the density of planetesimals is diminished by ejection \citep{b7,b9} and the planetesimal migration caused by the gas drag effect collaborate to empty this zone.   

Once the core became massive enough to retain a gas envelope, the effect of this atmospheric gas drag on the planetesimals increases the collision cross section of the protoplanet. This process was also taken into account in the model.

When the core reaches the critical mass, the gas accretion process begins. In this work only those embryos with very few gas are considered, because the process of collisions between gas giant is poorly understood. For this reason we considered only those embryos with masses $M_t<10M_{\oplus}$. Nevertheless I will explain the gas accretion model considered for those protoplanets which attain the critical mass necessary to start the gas accretion process before reaching the $10M_{\oplus}$. 

We assume that the critical mass necessary to start the gas accretion process is given by 

\begin{equation}
M_{crit}\sim \big(\frac{\dot{M_c}}{10^{-6}M_{\oplus}yr^{-1}}\big)^{\frac{1}{4}}
\end{equation}
This process occurs on a rate,

\begin{equation} \label{acregas1}
\frac{dM_g}{dt}=\frac{M_t}{\tau_g}
\end{equation}
where $M_g$ is the mass of the surrounding envelope and $\tau_g$ is its characteristic growth time, 

\begin{equation} \label{acregas2}
\tau_g=1.64x10^9 \bigg(\frac{M_t}{M_{\oplus}}\bigg)^{-1.91}yrs
\end{equation}
this values were fitted from results obtained by \citet{b10} as is explained in \citet{b1}.

\subsection{Angular momentum transfer due to the accretion of planetesimals}\label{planetesimales}

Our model also includes the acquisition of spin angular momentum by the growing embryos due to the accretion of mass in the form of planetesimals. Mutual impacts between embryos contribute to the stochastic component of the angular momentum. On the other hand, accretion of a large number of small planetesimals produces an ordered spin angular momentum, which will be discuss in this section. 

In order to model the angular momentum accreted by the protoplanets due to the planetesimal mass accretion, we follow the work of \citet{b34}. Their model depends on two parameters: 
\begin{itemize} 
\item The  relevance  of  the velocity  dispersion of the planetesimals in the planet's neighbourhood respect to the differential rotation  of the  planetesimal disc,  and
\item the  importance of  the planet's gravity as compared to the self gravity of the disc.
\end{itemize}
In the oligarchic growth regime, it is straightforward to demonstrate that the appropriate regime is that of high dispersion and strong gravity \citep{b34}. In this case, if we analyse the contribution of the small planetesimals, we would found that the stochastic component is near one order of magnitude smaller that the ordered one. Therefore, we add to our model, only the ordered accretion of angular momentum due to the planetesimal accretion.

According to the appropriate three dimensional case of \citet{b34}, the $z$ component of the angular momentum ${\bf L}$ due only to the ordered component is given by

\begin{equation}\label{lz}
L_{z,ord}^2\simeq M_t^2\, \Omega^2 \, R_p^4 \bigg(\frac{9}{7^2}\lambda\bigg)
\end{equation}
where

\begin{equation}\label{lamda}
\lambda=\frac{R_H^3\Omega^2}{R_p\sigma^2} \qquad R_H=a\bigg(\frac{M_t}{M_{\star}}\bigg)^{1/3}
\end{equation}

and the velocity dispersion is

\begin{equation}\label{sigma}
\sigma^2=\frac{1}{2}a^2\Omega^2e_m^2
\end{equation}

We assume that the RSM eccentricity of planetesimals in the disc is the equilibrium value found by \citet{b7}, which is given by equation \ref{em}. Introducing equations \ref{lamda}, \ref{sigma} and \ref{em} in equation \ref{lz}, we obtain the expression for the $L_z$ component due to the accretion of planetesimals, 

\begin{equation}\label{lz2}
L_{z,ord}\simeq 0.462 \frac{M_t^{5/3}\Omega a^{7/5} \rho_{gas}^{2/5}}{M_{\star}^{1/3}m^{2/15}\rho^{3/5}}
\end{equation}
Where all the units must be in cgs and the $L_{z,ord}$ is in $g\,cm^2 \,s^{-1}$.
Then at each  time step, the $z$ component of the angular momentum ${\bf L}$ changes by an amount

\begin{equation}
\Delta L_{z,ord}\simeq 6.54x10^{-7} \frac{\rho_{gas}^{2/5}\, M_{\star}^{1/6}\, M_t^{2/3}}{\rho^{11/15}\, a^{1/10}}\Delta M_t
\end{equation}

where $\Delta M_t$ is the mass accreted in the form of planetesimals during the given time step. All the units are in cgs and the $\Delta L_{z,ord}$ is in $g\,cm^2 \,s^{-1}$.

In order to obtain an estimate of which is the angular momentum acquired by a planet due only to the ordered component, we calculated the dependence of $L_{z,ord}$ with the embryo's mass for a protoplanet located at $1AU$ from a star like the Sun and we found,

\begin{equation}\label{lz1ua}
L_{z,ord}(1AU)=1.34x10^{40}\bigg(\frac{M_t}{1M_{\oplus}}\bigg)^{5/3}\, g\,cm^2 \, s^{-1}  
\end{equation}

Using equation \ref{lz2}, we can obtain the rotation period reach by an embryo which only acquire angular momentum due to the accretion of planetesimals, which is,

\begin{displaymath}
P_{ord}\simeq 150 \bigg(\frac{\rho}{3g/cm^3}\bigg)^{1/15}\bigg(\frac{a}{1au}\bigg)^{1/10}\bigg(\frac{\rho_{gas}}{\rho_{gas}(1au)}\bigg)^{-2/5}
\end{displaymath}
\begin{equation}\label{pord}
\bigg(\frac{M_{\star}}{M_{\odot}}\bigg)^{-1/6}hours
\end{equation}
for a planet located at $1AU$ and which orbits a star with $1M_{\odot}$ is $P_{ord}\simeq 150 \,  hours$.  

Nevertheless this is not the only mechanism able to change the spin of the emerging embryos, the collisions between the protoplanets have a huge importance in order to determine the final spin parameters.

\subsection{Collisions}\label{choques}

In the later stages of planetary formation, collisions represent an important evolutionary process which plays a significant role in determining the final mass and spin state of the planets. These interactions are not fully understood, here we explain the model considered in the work, which is very simple but enables us to get some conclusions regarding the primordial obliquities and rotation periods of planets.  
          
When two protoplanets are too close to each other, mutual gravitational influence can pump up their eccentricities to values sufficient to ensure their orbits to cross. Once the protoplanets have perturbed one another into crossing orbits, their subsequent orbital evolution is governed by close gravitational encounters and violent, highly inelastic collisions.

Under the assumption of perfect accretion in collisions, we consider that a merger between protoplanets will occur if their orbital spacing, $\Delta a$, is less than 3.5 Hill radius. 

The magnitude of the relative velocity at which two bodies of total masses $M_{t,1}$ and $M_{t,2}$ and radii $R_1$ and $R_2$ collide is

\begin{equation} 
v_{col}=(v_{rel}^2+v_e^2)^{\frac{1}{2}}
\end{equation}
where $v_{rel}$ is the relative velocity between the two bodies far form an encounter and $v_e$ is the scape velocity from the point of contact, given by

\begin{equation}
v_e=\Big(2G\frac{M_{t,1}+M_{t,2}}{R_1+R_2}\Big)^{\frac{1}{2}}
\end{equation}

The relative velocity between two embryos of orbital velocities $v_1$ and $v_2$ is 

\begin{displaymath}
\vec v_{rel}=\vec v_1-\vec v_2
\end{displaymath}

Considering that $a_2=a_1+\Delta a$, with $\Delta a<<a_1$ and that the collisions are randomly oriented  we obtained the following equation which shows the relative velocity between the embryos \citep{b8},

\begin{equation} 
v_{rel}\simeq \Omega \frac{\Delta a}{2} 
\end{equation}
with $\Omega$ the orbital angular velocity (to our degree of approximation
it is equivalent to adopt $a=a_1$ or $a_2$, but we choose as $a$ the semi major axis of the more massive planet). The distribution of velocities is isotropic, so the direction is chosen randomly with an isotropic probability distribution. 

We assume that in the beginning the embryos do not rotate but during its evolution they acquire spin angular momentum by the accretion of planetesimals (as seen in section \ref{planetesimales}) and by the collisions with other embryos. Here we analyse the total spin angular momentum of the resultant embryo acquired after a collision which is,

\begin{equation} 
\vec L_{imp}=\vec L_{col}+\vec L_{spin}
\end{equation}
with $\vec L_{spin}$ the sum of the spin of the target, $\vec L_{spin,tar}$, and the spin of the impactor, $\vec L_{spin,im}$ and $L_{col}$ is the spin angular momentum delivered by the impactor during a collision where the impact point on the surface of the target is randomly calculated by assuming spherical embryos.  

Our assumption of perfect accretion occasionally allows particles to spin faster than break-up and destroy the embryo. This happen when the acceleration produced by the rotation are higher than gravity, which means,

\begin{equation}
R \omega^2 > \frac{G M_{t}}{R^2}
\end{equation}  
with $\omega$ the rotation angular velocity. This condition leads to a critical value for the angular velocity,

\begin{equation}\label{wcrit}
\omega_{crit}= \Big(\frac{G M_t}{R^3}\Big)^{\frac{1}{2}}
\end{equation}
beyond which the embryo is gravitationally unbound.

\subsection{Orbital evolution}

When a protoplanet is embedded in a disc, their interaction are significant and lead to different regimes of planetary migration regarding the embryo mass. When the protoplanet involved is a low-mass planet, the interaction can be calculated using a linear theory which leads to a type I planetary migration, but when the planet reaches the mass necessary to open up a gap in its orbit, the disc response can no longer be treated as linear and it leads to the type II regime. The critical mass is derived from the condition 

\begin{equation}
 r_H \geq h
\end{equation}
which is necessary for open a gap \citep{b12}, and where h is the disc scale of height.

The model for type I and II migration is the same considered before \citep{b2}, which is essentially the same used by \citet{b11} in their model, where the time-scales are given by

 \begin{equation}\label{migI}
\tau_{migI}= -\frac{a}{\dot{a}}\simeq 1.26x10^5\frac{1}{C_{migI}}\frac{1}{f_g}\Big(\frac{M_p}{M_{\oplus}}\Big)^{-1}\big(\frac{a}{1au}\big)^{\frac{3}{2}}\Big(\frac{M_{\star}}{M_{\oplus}}\Big)^{\frac{3}{2}}yrs
\end{equation}
\begin{equation}
\tau_{migII}=0.8x 10^6 f_g^{-1}\Big(\frac{M_p}{M_J}\Big)\Big(\frac{M_{\odot}}{M_{\star}}\Big)\Big(\frac{\alpha}{10^{-4}}\Big)^{-1}\Big(\frac{a}{1au}\Big)^{\frac{1}{2}}yrs
\end{equation}  
 with $\alpha=10^{-3}$ a dimensionless parameter which characterises the viscosity and the factor $\frac{1}{C_{migI}}$ is introduced for considering other important effects that might slow down the migration, without introducing a mayor degree of complexity to the model.

 We assume that both migration mechanisms stop when the core reaches the inner edge of the disc.

\section{Results}

We investigate the statistical properties of primordial planetary spin resulting from the process of planetary formation through numerical simulations. In this section we show our main results. 

\subsection{Some statistics of planets found}

We generate 1000 discs, for every system the mass of the star is taken random from values which follow a uniform log distribution in the range of $0.7-1.4M_{\odot}$, and the time-scale for the depletion of the disc of gas has a uniform distribution in log scale between $10^6$ and $10^7$ years. Each system evolves for $2x10^7$ years.

We consider the formation of planetary systems which have suffered type I and II regimes of planetary migration, where the retardation constant for type I regime of migration is taken as $C_{mgiI}=0.1$.

\begin{figure}
  \begin{center}
    \includegraphics[angle=270,width=.44\textwidth]{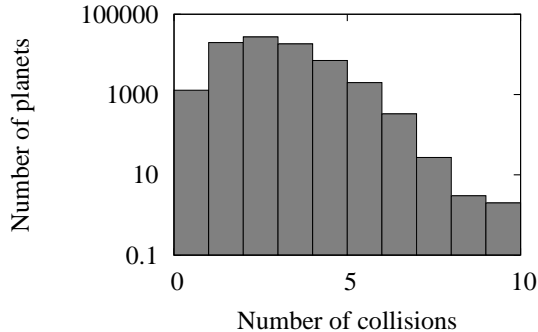}
  \end{center}
  \caption{Histogram of the number of collisions suffered by each embryo at the end of the simulation. We see that a few collisions determine the primordial planetary spin.}
  \label{fig:ncol}
\end{figure}

Figure \ref{fig:ncol} shows an histogram of the number of collisions suffered by each embryo over the $2x10^7$ years. We note that most of the planets suffer less than 5 impacts during its formation, which means that in most of the cases primordial spins of planets are randomly determined by a very few impacts suffered during accretion. On the other hand we also found some planets which have more impacts. This is due to the migration of the embryos which makes some of them move rapidly towards the star and suffer more collisions than the most external ones, which have very few embryos to collide with. 

\begin{figure}
  \begin{center}
    \includegraphics[angle=270,width=.44\textwidth]{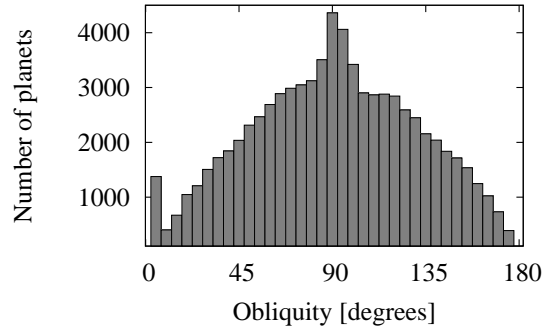}
  \end{center}
  \caption{Histogram of primordial obliquities found, which correspond to an isotropic distribution.}
  \label{fig:distriobli}
\end{figure}

We also analyse the distribution of planetary primordial obliquities. Figure \ref{fig:distriobli} shows this distribution where we see that the obliquity distribution corresponds to an isotropic distribution of the spin vector, given by 

\begin{equation}
 p(\epsilon)=\frac{1}{2} sin(\epsilon)
\end{equation}
this result confirms the earlier findings of \citet{b13,b19} and \citet{b14}, which were obtained using N-body simulations, and is due to the fact that during this stage of planetary formation, the scale of height of the disc is much larger than the size of the embryos, so collisions can occur in any direction. For this reason, the result is indeed independent of the orbital evolution of the embryo: the isotropic distribution is maintained if we consider planetary migration or if we do not, the only difference is the amount of collisions suffer by the embryos. 

Since the ordered component has obliquity of 0 or 180 degrees, we would expect that this will have consequences in the obliquities distribution, but the effect of the ordered accretion is important in those embryos which do not suffer any collision during their growth. In these cases, the angular momentum acquired by the accretion of planetesimals determines their final spin state. On the other hand for those embryos that suffer great collisions during their formation, their obliquities are determined by the momentum acquired during the impacts because this stochastic component is very strong and dominates the final state of the embryo. According to our results, most of the embryos suffer collisions during their formation and for this reason we do not found significant changes in the obliquity statistics made.

We also note that there are a large amount of embryos that did not collide with any other. These are tiny embryos who grew in a low mass disc and are located near to the inner edge of the disc, as a consequence, they were not eaten by a larger embryo migrating towards the star. 

\begin{figure}
  \begin{center}
    \includegraphics[angle=270,width=.44\textwidth]{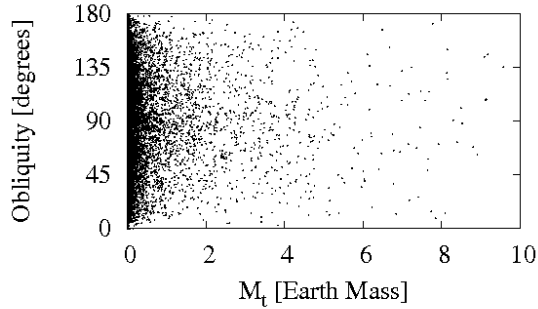}
  \end{center}
  \caption{Obliquity of the surviving planets plotted against their mass.}
  \label{fig:m-o}
\end{figure}

We show in figure \ref{fig:m-o} the obliquity of the planets plot against their mass. This plot shows that we can found, with equal probability, terrestrial planets with obliquities between $0$ and $180$ $degrees$, which means planets who rotate in a direct or indirect sense, independently of its mass. This result tells us that the primordial spins of planets are not those commonly observed in the terrestrial planets in our own Solar System, whose current spin axes are more or less perpendicular to their orbital planes (except for Venus). However, the spin axes of the terrestrial planets are not primordial, so this does not necessarily indicate a problem in the model consider here. Other studies such as planet-host star or planet-satellite's tidal interaction \citet{b18,b17}, among others, must be taken into account for explaining the present obliquities of the terrestrial planets.

\subsubsection{The study of the rotation periods}

\begin{figure}
  \begin{center}
 \includegraphics[angle=270,width=.44\textwidth]{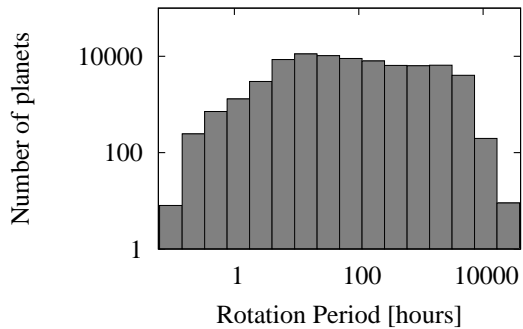}
\end{center}
  \caption{Distribution of rotation periods of all the planets found in our simulations.}
  \label{fig:distriperrot}
\end{figure}

The rotation periods of planets were calculated assuming that the protoplanets were spheres of uniform density and the distribution found is shown in figure \ref{fig:distriperrot}, where we note that most of the planets reach rotation periods larger than $\sim 10$$hours$ but there are also a large amount of planets with periods between $0.1$ and $10$$hours$. The planets who reach spin periods less than $0.1$ $hours$ are really rare, because at that rotation periods they use to have a angular velocities larger than the critical one.

\begin{figure}
  \begin{center}
 \includegraphics[angle=270,width=.48\textwidth]{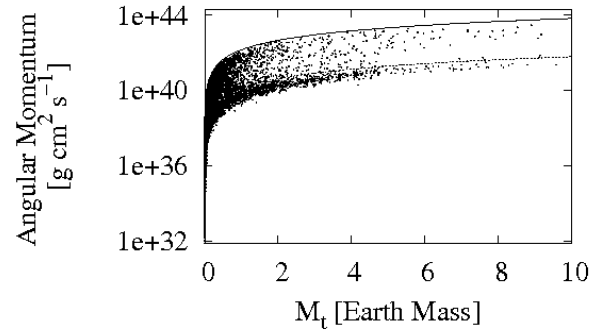}
\end{center}
  \caption{Mass and angular momentum of the planets found in our simulations. The solid line represents the $L_{crit}$ beyond which the embryos are gravitationally unbound. The dotted line represent the value for the ordered angular momentum reached by the cores located at $1AU$ from the sun (see equation $\ref{lz1ua}$).}  
  \label{fig:momento-masa}
\end{figure}

In order to understand the rotation periods distribution, we study the angular momentum of the planets formed. Figure \ref{fig:momento-masa} shows the angular momentum as a function of the mass of the embryos, where the solid line represent the $L_{crit}$ beyond which the planets are gravitationally unbound, which can be deduced from equation \ref{wcrit}. The dotted line shows the angular momentum due to the ordered component that acquire a planet located at $1 AU$ from the Sun, which was deduced from equation \ref{lz2} and shown in equation \ref{lz1ua}. We note that those embryos who did not suffer any collision and acquire angular momentum only by the accretion of planetesimals, should have $L$ near the dotted line, while those who experienced the change of momentum due to one or more impacts, could reach a larger angular momentum but always below the stability limit. 

We know that the rotation period is inversely proportional to the angular momentum, so the higher the angular momentum, the shorter the period, as a consequence there can not be planets with small periods, and that is why we have an absence of planets with periods less than $\sim 0.5$ hours in the figure\ref{fig:distriperrot}.

We also study the evolution of the rotation periods in three different simulation times: at $1000$, $10^5$ and $2x10^7$ $years$, which is the final simulation time. 

\begin{figure}
  \begin{center}
    \subfigure[]{\label{fig:m-p-evo1}\includegraphics[angle=270,width=.46\textwidth]{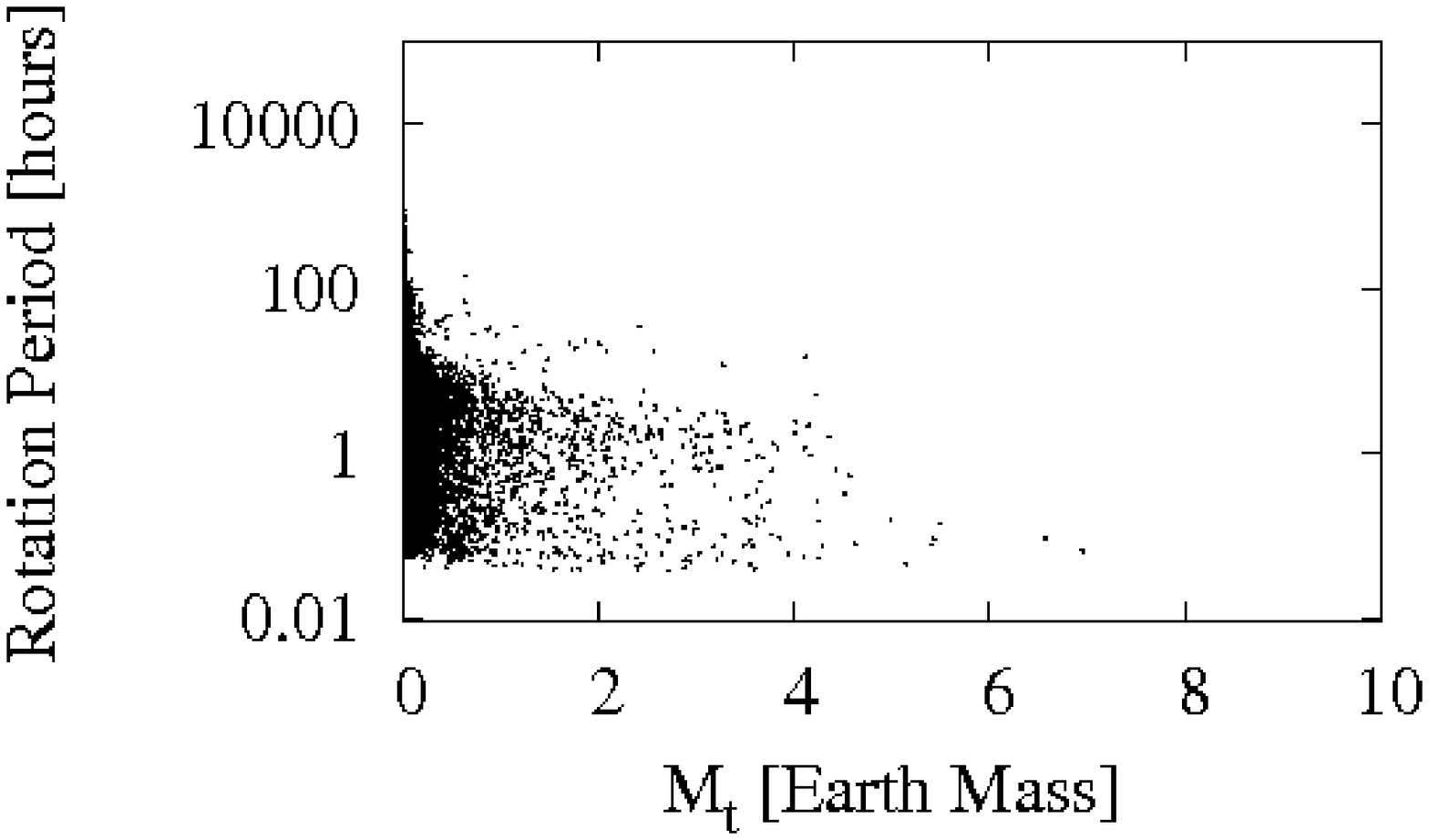}}
    \subfigure[]{\label{fig:m-p-evo2}\includegraphics[angle=270,width=.46\textwidth]{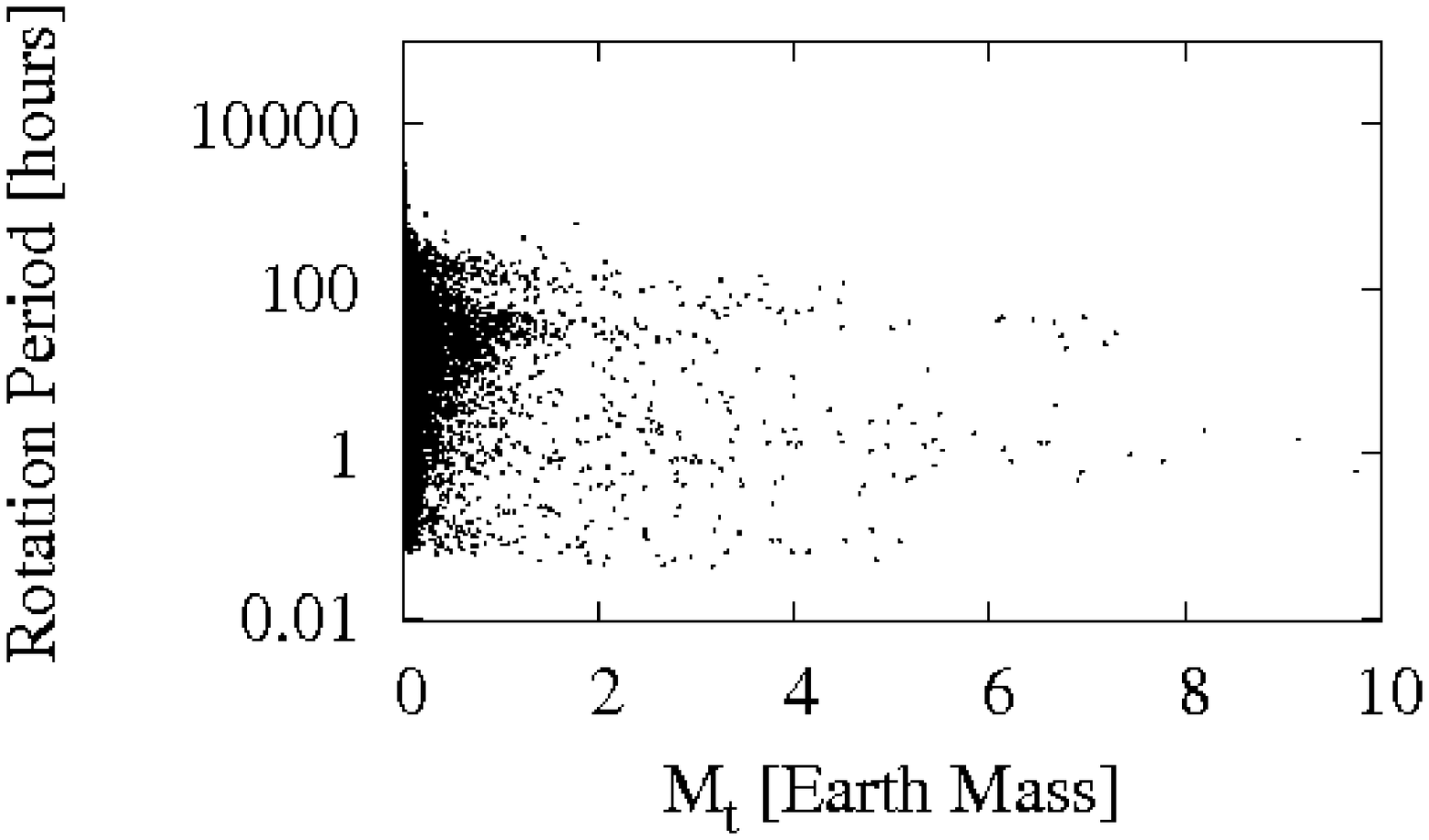}}
    \subfigure[]{\label{fig:m-p-evototal}\includegraphics[angle=270,width=.46\textwidth]{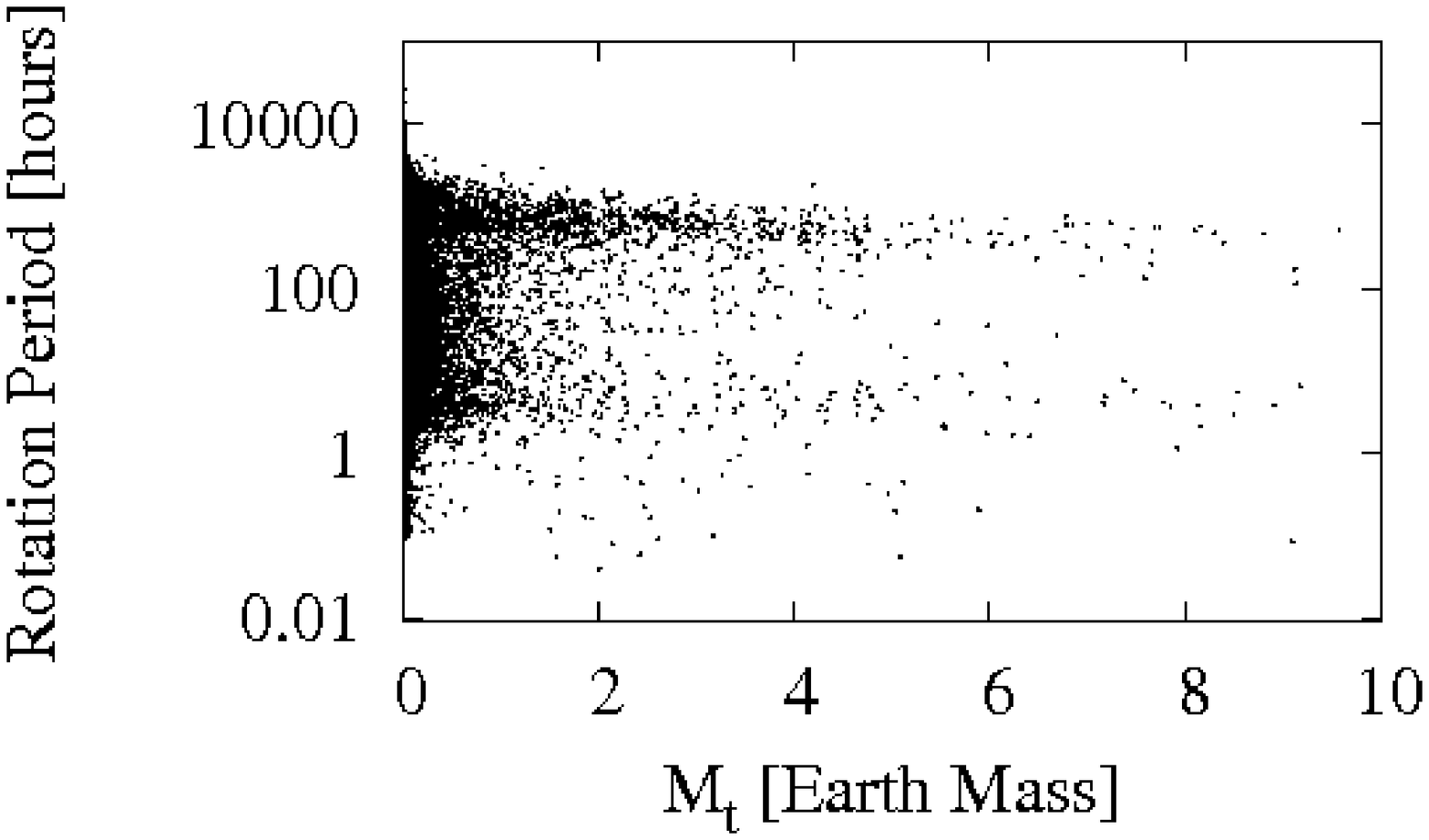}}
  \end{center}
  \caption{The rotation period is plot against the embryos' mass for different times. Figure \ref{fig:m-p-evo1} presents the results at $1000$ $years$, in figure \ref{fig:m-p-evo2} has pass $10^5$ $years$ and the last figure, \ref{fig:m-p-evototal} shows the distribution at the end of the simulation ($2x10^7$ $years$).}
  \label{fig:m-perrot}
\end{figure}

The rotation periods of embryos as a function of its mass in the three different times is plot in figure \ref{fig:m-perrot}.  In the first figure (\ref{fig:m-p-evo1}), when $1000$ $years$ have passed, we found only embryos with rotation periods until $100$ $hours$, and small masses, which is probably due to the short time that has passed, the embryos did not have much time to grow. We also observe a small population of more massive embryos with shorter periods, some of them with masses of up to $7M_{\oplus}$.  Since very few time has passed, they are probably embryos located in the interior region of the disc. This region is rich in solids and this favors the rapid growth of the embryos cores', which makes them the firsts to suffer a large amount of collisions and hence increase their spin angular velocities, leaving them to the brink of instability.

As time passes (figure \ref{fig:m-p-evo2}) we note that the embryos acquire larger periods, and the amount of embryos with small periods decreases. Finally, at the end of the simulation (figure \ref{fig:m-p-evototal}), we note a well-marked difference between the few planets with periods less than $\sim 1$ $hours$ and the rest of the population. These are very rare planets. As seen in equation \ref{wcrit}, those embryos with small rotation periods rotate rapidly, so their spin angular velocities are high enough to overcome the critical rotation angular velocity for rotation instability. As a consequence we find a small amount of planets with this periods, only a very few percent survive, and the surviving ones have mainly small masses. 

On the other hand we observe that most of the planets have reached rotation periods of up to $\sim 10000$ $hours$. These are probably the embryos that only acquire their angular momentum by the accretion of panetesimals.

\begin{figure}
  \begin{center}
   \includegraphics[angle=270,width=.48\textwidth]{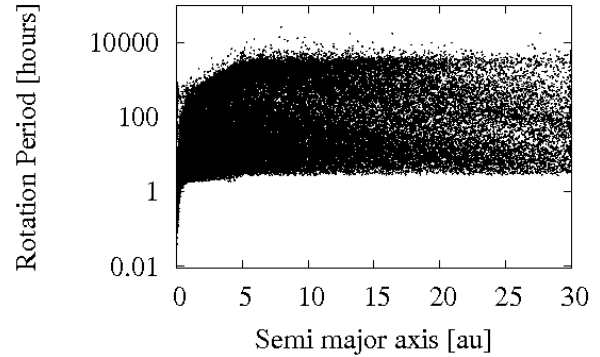}
  \end{center}
  \caption{The figure shows the rotation period and semi major axis of all the surviving planets at the end of the simulation.} 
  \label{fig:a-p-evotot}
\end{figure}

In figure \ref{fig:a-p-evotot} the rotation period is plot as a function of the embryo's semi major axis, where those planets with the largest rotation periods probably acquired them mainly by the accretion of planetesimals, while those with the shorter periods need one or more impacts for having that spin. 

We can also compare our results with those observed in the terrestrial planets in our own solar System. In the case of Mercury and Venus, whose current rotation periods are $58.64$ and $243.01 \; days$ respectively, their spin rates have undergone great changes since their formation and could not be considered as primordial. The close proximity of these planets to the Sun produces a tidal dissipation that has slowed down the spin rates, then the primordial values of the rotation period must have been much lower than those currently observed, for this reason we can not compare with these planets.      

The case of Mars and the Earth is different, because they are located far from the Sun and that is why Solar tides have not altered their spins appreciably. The rotation speed of Mars can be considered as primordial, because its satellites are so small that have not influenced appreciably in the spin rates' evolution. In the case of the Earth, while the magnitude of the Earth-Moon system' s angular momentum has been approximately constant since its formation, the Earth' spin has been slowed down by lunar, since tidal interactions have transferred angular momentum from the Earth to the Moon. 
 
While the current (and primordial) rotation period of Mars is $\sim 24.5 \, hours$, the one that should have had if all its mass were obtained only by the accretion of planetesimals (no collisions involved), could be deduced from equation \ref{pord} and is $247.9 \, hours$. In the case of the Earth, the maximum value reached by the rotation period since it was formed is the current one, and the one obtained only by the ordered component is $150 \, hours$. According to this results Mars and the Earth did not acquire their rotation periods only by the accretion of planetesimals, but during one or more impacts during its formation. 

We also note that in the population of planets shown in figure \ref{fig:a-p-evotot}, we found a large sample of planets with the characteristics of the Terrestrial Planets. 

\subsubsection{Mass and Semi major axis distribution}

In our previous works \citet{b1,b2}, we have studied the changes in the mass and semi major axis diagram due to different factors. As \citet{b9} have shown, this diagram shows the process of planetary formation, where different regimes of planetary growth were found depending on the material available in the region on the disc where it was formed. 

As equation \ref{core-accretion} shows, the cores' planetesimal accretion rate depend on the region where the embryo is located and the solids available, which are larger at the smallest ($a<1$ $au$) semi major axis (equation \ref{distrisolidos}). For this reason, we found a rapid cores' growth in the inner regions of the disc, where $a<1$ $au$, and the lowest solid accretion rates are found in the outer regions of the disc, where the embryos take longer to grow.

\begin{figure}
  \begin{center}
    \subfigure[]{\label{fig:m-a-evo1}\includegraphics[angle=270,width=.46\textwidth]{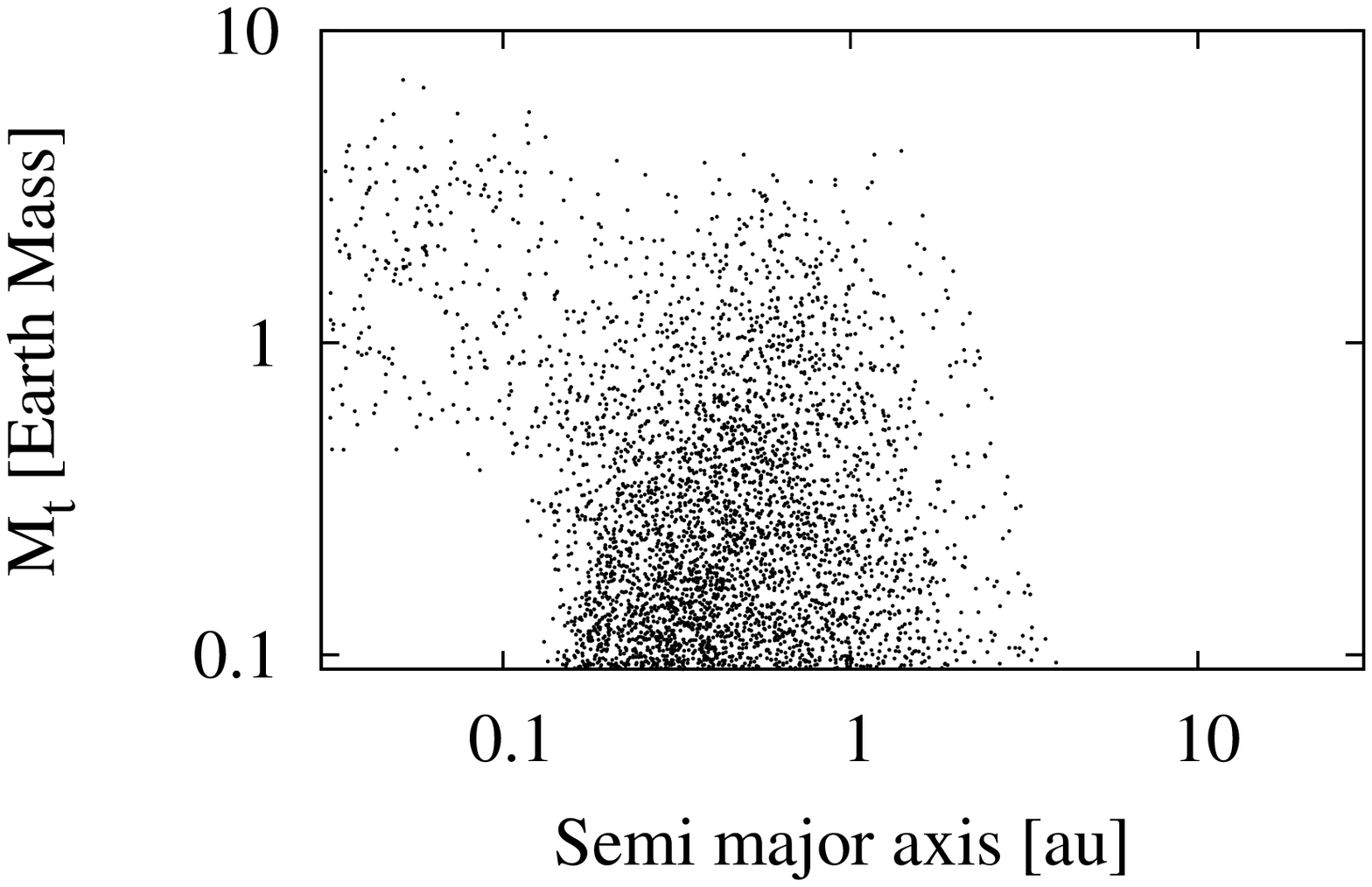}}
    \subfigure[]{\label{fig:m-a-evo2}\includegraphics[angle=270,width=.46\textwidth]{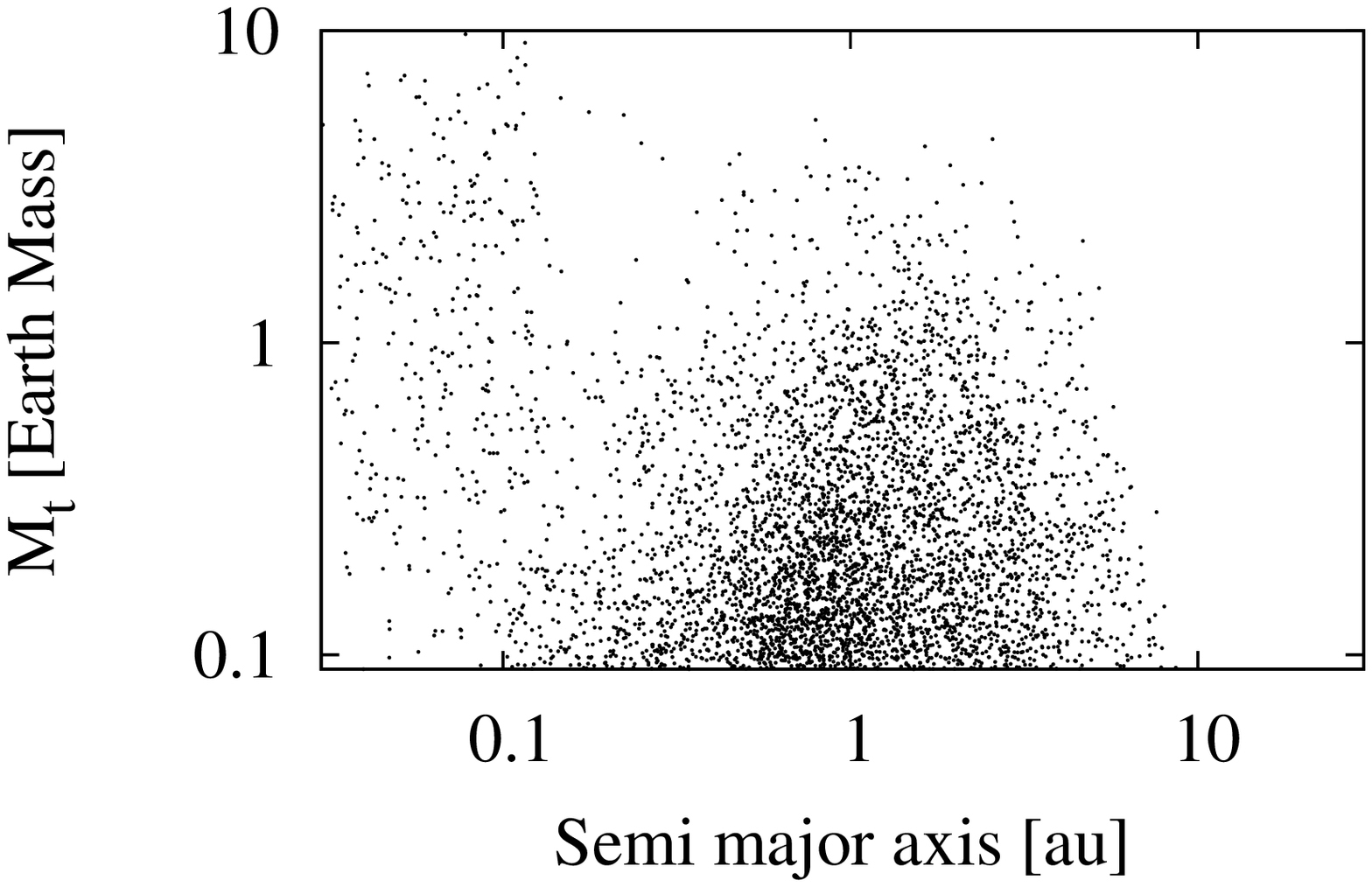}}
    \subfigure[]{\label{fig:m-a-evototal}\includegraphics[angle=270,width=.46\textwidth]{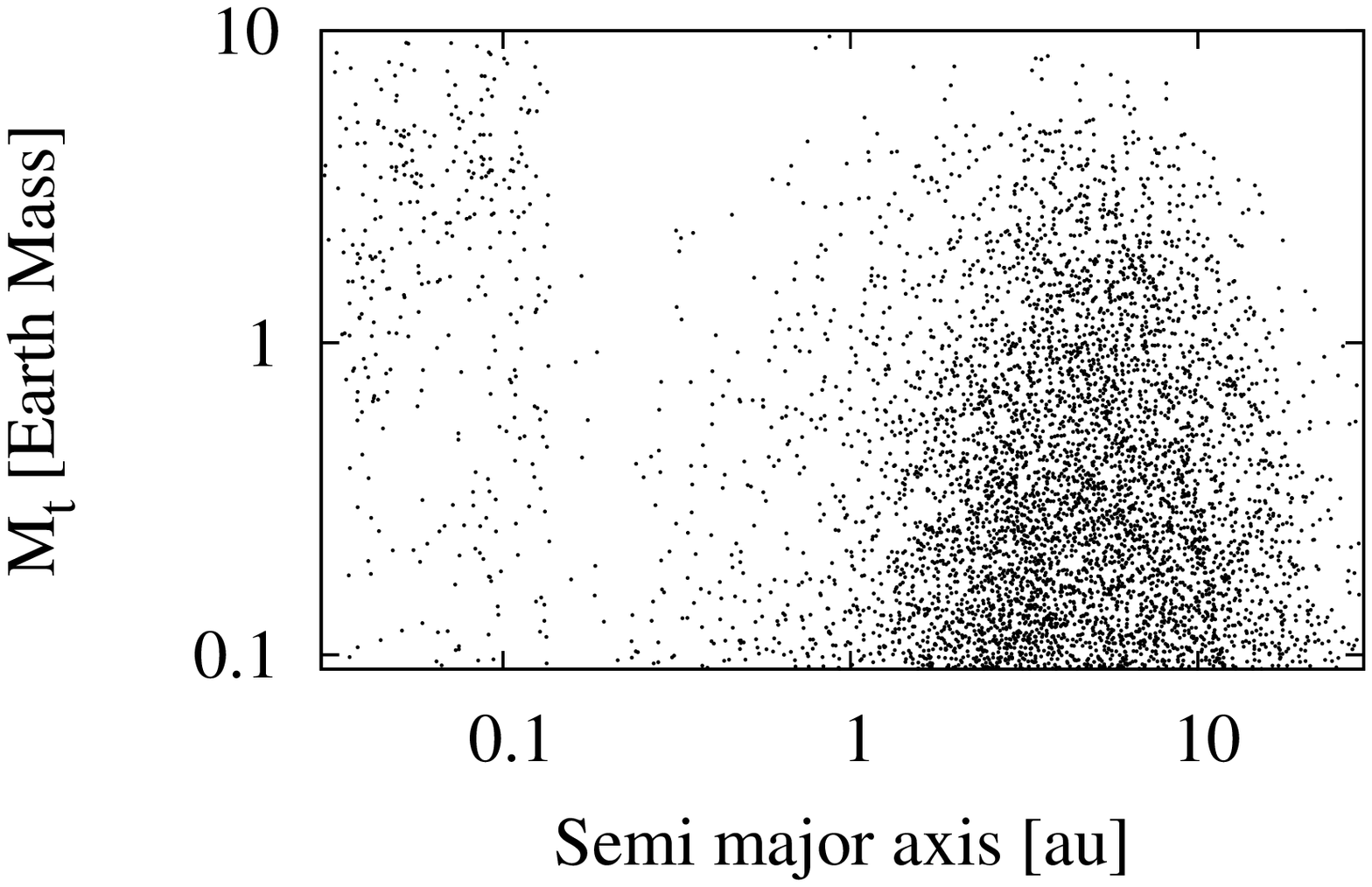}}
  \end{center}
  \caption{Mass and semi major axis distribution found in different evolution times. Figure \ref{fig:m-a-evo1} shows the results at $1000$ $years$, in figure \ref{fig:m-a-evo2} has passed $10^5$ $years$ and in figure \ref{fig:m-a-evototal} we plot the results at the end of the simulation. In all the simulations we consider the fragmentation of embryos by collisions.}
  \label{fig:a-perrot}
\end{figure}

As seen in previous section, the region in the disc where a planet grows, has also a strong influence in the rotation period reached by the planet, because a smaller semi major axis ensures that the planets have more solids available and they will accrete more angular momentum. Besides, if they are in a region with a large density of embryos, they would have a large probability of collisions, that will change their spin too. Figures \ref{fig:m-a-evo1},\ref{fig:m-a-evo2} and \ref{fig:m-a-evototal} show the mass and semi major axis distribution in the three times studied: $1000$, $10^5$ and $2x10^7$ $years$ respectively. As seen in the first figure, in the begging the embryos who grow faster are those initially located really close to the star, in the inner part of the disc and as seen in figure \ref{fig:a-p-evotot}, between these embryos we also found those with the smallest rotation periods. On the other hand, those embryos located in the intermediate and outer part of the disc remain almost with their initial mass. As time passes (figure \ref{fig:m-a-evo2}), we note that the embryos who grow faster have migrated closer to the star or fragmented and disappear, while the embryos located in the intermediate region began to grow. The results obtained at the end of the simulation, represented in figure \ref{fig:m-a-evototal}, show that those embryos who were located initially in the intermediate region of the disc, have migrated to the star (or fragmented and disappear) and those in the outer regions of the disc, have now starting to grow reaching rotation periods until $\sim 10000 \; hours$. 

\begin{figure}
  \begin{center}
  \includegraphics[angle=270,width=.46\textwidth]{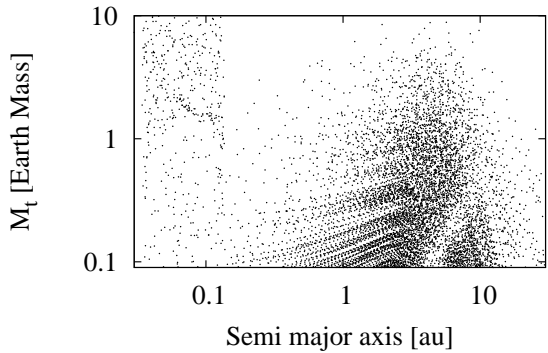}
\end{center}
  \caption{Mass and semi major axis distribution found without considering the fragmentation of planets by collisions.}
  \label{fig:m-a-evototal-antes}
\end{figure}

In our previous works we have focused our attention to those planets with masses larger than $1 M_{\oplus}$. Here we study planets with masses less than $10 M_{\oplus}$ and found another important effect which changes the distribution of mass and semi major axis of terrestrial planets. As seen in before, embryos with small periods, can not resist the acceleration due to rotation and destroyed themselves. This fact will change slightly the mass and semi major axis distribution obtained in our previous works. With the aim of comparing with our previous results, we plot in figure \ref{fig:m-a-evototal-antes}, the mass and semi major axis distribution found when fragmentation by collisions was not taken into account.

Comparing figure \ref{fig:m-a-evototal} with figure \ref{fig:m-a-evototal-antes}, we observe fewer planets considering the fragmentation by collisions that those found in the other case. So this is an important effect that must be considered when working with terrestrial planets.

\subsection{Those who did not survive}

As we have shown in section \ref{choques}, there is a stability limit beyond which the planets are not able to remain united and disarmed. Here we show some statistics regarding those planets who could not survive.

\begin{figure}
  \begin{center}
    \subfigure[]{\label{fig:m-p-evo1-rotos}\includegraphics[angle=270,width=.46\textwidth]{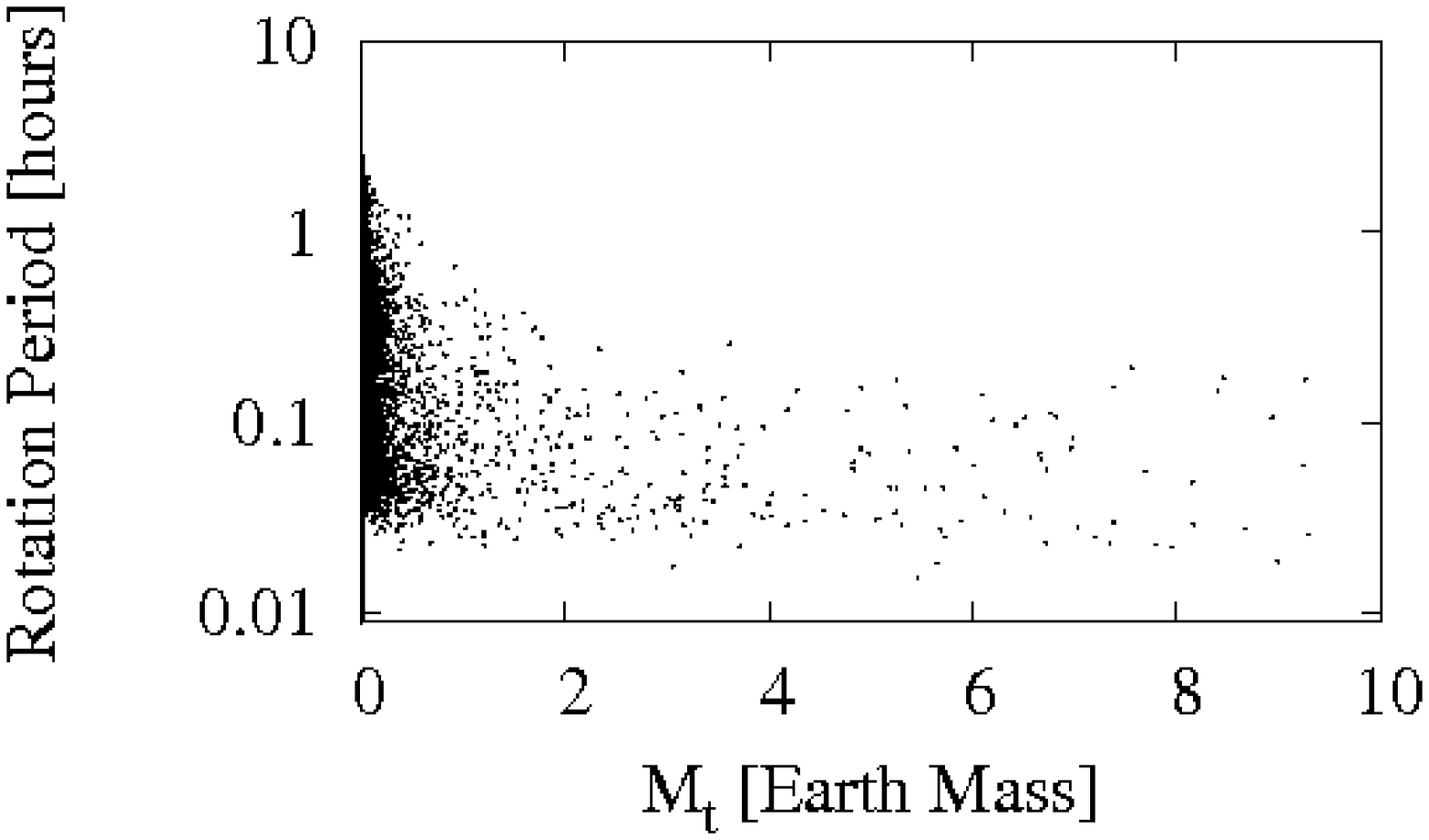}}
    \subfigure[]{\label{fig:m-p-evo2-rotos}\includegraphics[angle=270,width=.46\textwidth]{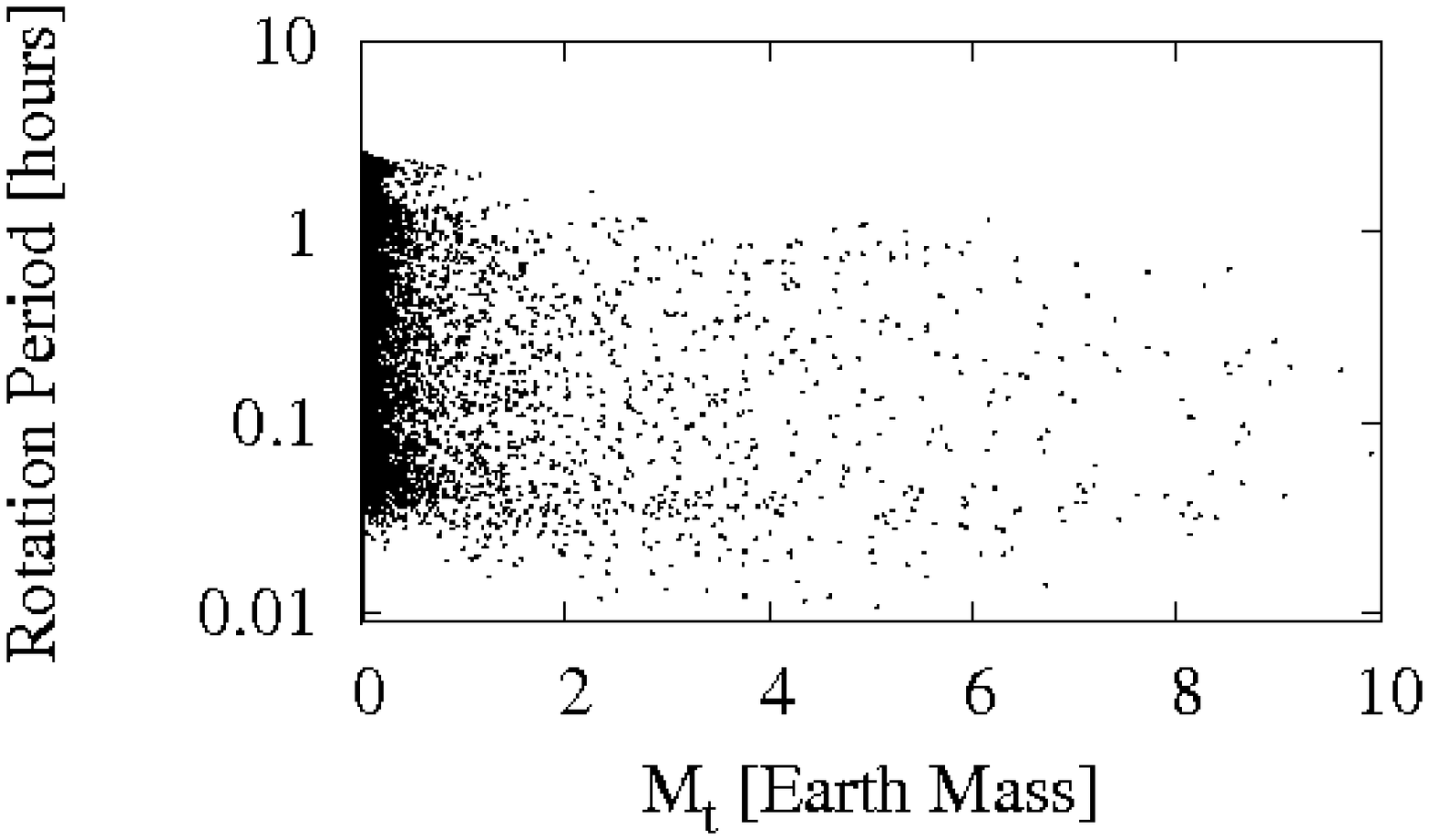}}
    \subfigure[]{\label{fig:m-p-evototal-rotos}\includegraphics[angle=270,width=.46\textwidth]{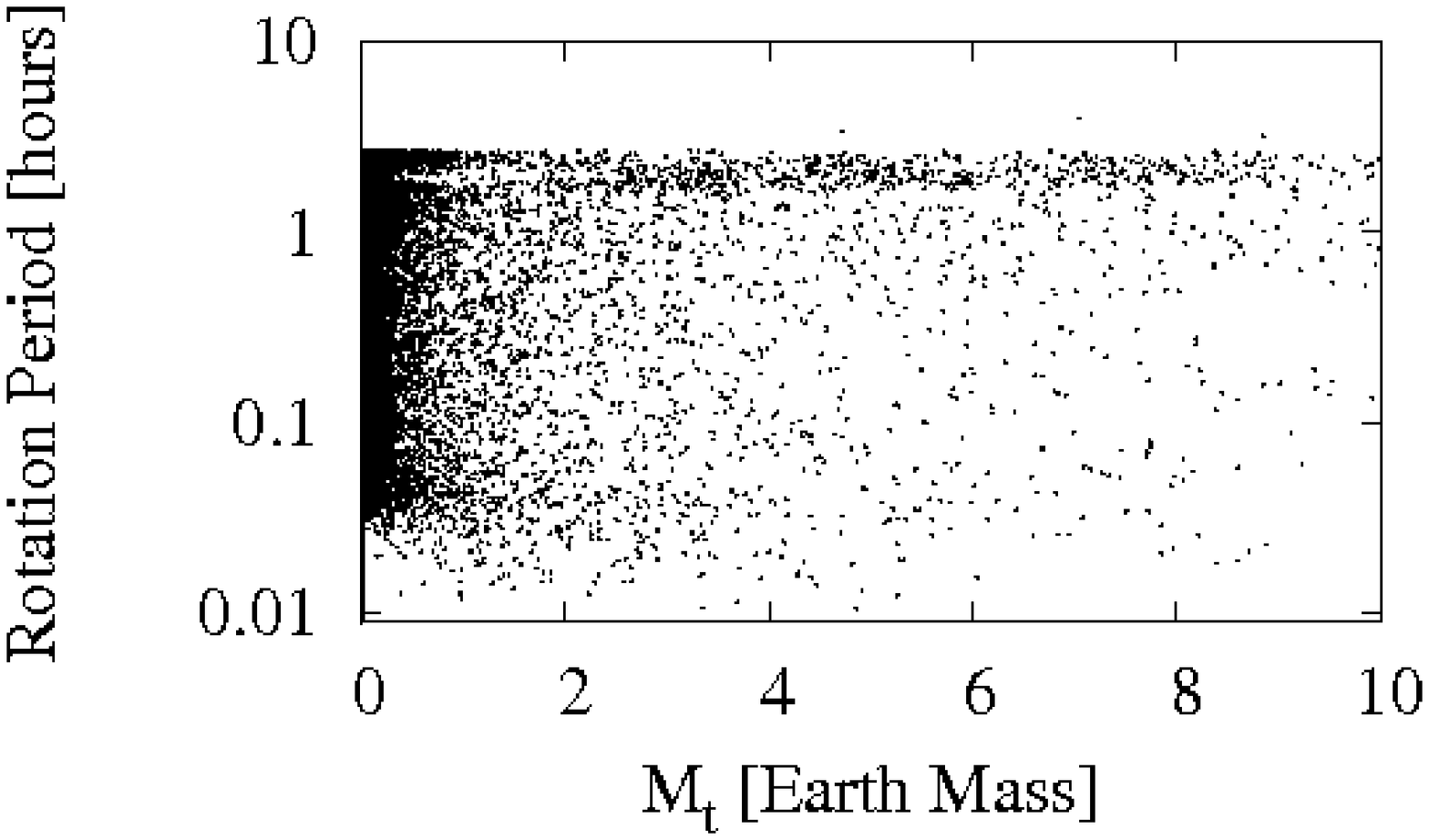}}
  \end{center}
  \caption{Mass and rotation period of embryos fragmented. Figure \ref{fig:m-p-evo1-rotos} shows those embryos fragmented before the firsts $1000$ $years$, in figure \ref{fig:m-p-evo2-rotos} we see the results at $10^5$ $years$ of simulation and figure \ref{fig:m-p-evototal-rotos} shows the total embryos who do not survive.}
  \label{fig:m-p-rotos}
\end{figure}

Figure \ref{fig:m-p-rotos} shows the mass and rotation period of these ``broken'' embryos as they were when reached the critical rotation angular velocity. Figure \ref{fig:m-p-evo1-rotos} presents the results of those embryos fragmented before the firsts $1000$ $years$, who are mostly those with periods less than $0.1 \; hours$, in figure \ref{fig:m-p-evo2-rotos} we see the results at $10^5$ $years$ and finally the last figure (figure \ref{fig:m-p-evototal-rotos}) shows the total embryos who do not survive. As seen in the figures all the broken embryos have rotation periods less than $\sim 2 \; hours$, which is approximately the critical period. We also observe those embryos with small spin periods are the firsts fragmented, then the embryos with periods near to $1 \; hours$, and finally those of $\sim 2 \; hours$ exceed the limit of stability. 

\section{Summary and conclusions}

In the process of planetary formation protoplanets collide with one another to form planets. We have investigated the final assemblage of terrestrial planets from protoplanets using a simple model which consider the oligarchic growth regime of protoplanets as initial condition in a disc where several embryos are allowed to form. As explained in our previous work the formation of several cores simultaneously in the disc has a strong influence on the dynamic of the planetesimal disc, which influences directly the growth of the embryos' cores and the final assemblage of planets found. 

In our model we also have included the interaction between the protoplanets and the disc, which leads to a planetary migration. When a embryo is migrating towards the central star it could perturbate the cores placed in its path, causing the accretion of the core in most cases, this collisions affect the spin state of the embryos.  

As collision among giant planets are poorly understood we have focused our attention on planets with masses less than $10 M_{\oplus}$ where a very simple model for planetary impacts has been considered. We suppose that when two embryos are a distance less than $3.5$ $R_{H}$ the merger between both protoplanets occurs, which leads to the union of two embryos to form a single body. This perfect accretion model produces spin rates that are too high and when the acceleration produced by the rotation is greater than those of gravity the body overcome the critical spin angular velocity for rotational instability and is fragmented. This simple model allows us to obtain some interesting results regarding the final properties of terrestrial planets. 

We also have considered the acquisition of angular momentum due to accretion of planetesimals. The accretion of a large amount of planetsimals produces an ordered spin that adds angular momentum to that acquired during collisions, so the final spin of the planets is a result of this two effects.

In order to analyse the statistical properties of the assembled planets we take different initial planetary system parameters, considering $1000$ different discs, where each planetary system evolves $2x10^7$ $years$. 

As in our previous works we have analysed the information provided by the mass and semi major axis diagram, which reflects the process of planetary formation. We observe fewer planets with masses less than $1 M_{\oplus}$ considering the fragmentation by collisions that those found in the without this effect. This means that the effect of fragmentation by collision has a strong influence on the final population of terrestrial planets formed and should be considered when these planets are involved.

We also have studied the effects produced by the collisions between the embryos, where we find that most of the planets suffer less than 5 impacts during its formation, which means that in most of the cases primordial spins of planets are randomly determined by a very few impacts suffered during accretion.

We also take special attention to final spin state, which means planetary obliquities and rotation periods, where we found that the distribution of obliquities of final planets is well expressed by an isotropic distribution, result that confirms those obtained previously by other authors \citet{b13,b14} and is independent on the planetary mass. This fact is in marked contrast to the terrestrial planets in our own Solar System, whose current spin axes are more or less perpendicular to their orbital planes (except for Venus). However, the spin axis of the terrestrial planets strongly depends on the gravitational perturbations from the other planets of the Solar System that create a large chaotic zone for their obliquities. So all of the terrestrial planets could have experienced large, chaotic variations in obliquity in their history, and this is why their obliquities can not be considered as primordial \citep{b36}. So the fact that the terrestrial planets in our Solar System present obliquities $\sim 0^{\circ}$ does not necessarily indicate a problem with the model considered here. Other studies such as body and atmospheric tides and core-mantle friction among others, must be taken into account for explaining the present obliquities of the terrestrial planets.  

Regarding the findings on the rotation period, we found that the primordial rotation periods of terrestrial planets are dependent on the semi major axis, which means on the region where the embryos were formed and evolved. 

On the one hand we note a very small population of planets with small rotation periods (less than $\sim 0.5 \; hours$), which are very rare planets, because at that rotation periods the spin angular velocities are high enough to overcome the critical rotation angular velocity for rotation instability. 

On the other hand there are a large population of embryos with rotation periods until $\simeq 10000 \; hours$. These planets with large rotation periods probably acquired them mainly by the accretion of planetesimals, while those with shorter periods need one or more impacts for acquire that spin. 

Another important result is that we have found a large population of planets with the characteristics of the Terrestrial Planets, and our results suggest that they did not acquire their rotation period only by the accretion of planetesimals, but during one or more impacts during their formation.

\end{document}